\documentclass[%
 reprint,
nofootinbib,
 amsmath,amssymb,
 aps,
prl
]{revtex4-2}

\usepackage{graphicx}
\usepackage{dcolumn}
\usepackage{bm}
\usepackage{xcolor}
\usepackage{hyperref}
\usepackage{amsmath}
\usepackage{slashed}
\usepackage{cleveref}

\begin{document}

\preprint{APS/123-QED}

\title{Could We Observe an Exploding Black Hole in the Near Future?}

\author{Michael J.~Baker}
\email{mjbaker@umass.edu}
\author{Joaquim Iguaz Juan}
\email{jiguazjuan@umass.edu}
\author{Aidan Symons}
\email{asymons@umass.edu}
\author{Andrea Thamm}
\email{athamm@umass.edu}
\affiliation{Department of Physics, University of Massachusetts, Amherst, MA 01003, USA
}%

\date{\today}

\begin{abstract}
Observation of gamma rays from an exploding black hole would provide strong evidence for primordial black holes, the first direct evidence of Hawking radiation, and definitive information on the particles present in nature.  However, indirect constraints suggest that direct observation of an exploding Schwarzschild black hole  at current or upcoming gamma ray observatories is implausible.  We introduce a dark-QED toy model consisting of a dark photon and a heavy dark electron. In this scenario a population of light primordial black holes charged under the dark $u(1)$ symmetry can become quasi-extremal, so they survive much longer than if they were uncharged, before discharging and exhibiting a Schwarzschild-like final explosion. We show that the answer is ``yes'', in this scenario the probability of observing an exploding black hole over the next $10$ years could potentially be over $90\%$.
\end{abstract}

\maketitle


\section{Introduction}

Following the first detection of a binary black hole merger by LIGO in 2015~\cite{LIGOScientific:2016aoc}, interest in the observational aspects of black holes has surged.  Beyond the stellar mass black holes observed via gravitational waves and the supermassive black holes that have been directly imaged~\cite{EventHorizonTelescope:2019dse}, there has also been a strong interest in Primordial Black Holes (PBHs).  PBHs are black holes that might have formed less than a second after the big bang~\cite{Hawking:1982ga, Carr:1975qj, Kawasaki:1997ju, Khlopov:2008qy, Brandenberger:2021zvn, Cotner:2018vug, Baker:2021nyl, Baker:2021sno}. 
Depending on their mass, PBHs have, among other applications, been invoked to constitute dark matter~\cite{Carr:2021bzv,Green:2020jor}, 
create dark matter particles when they evaporate~\cite{Morrison:2018xla,Hooper:2019gtx,Masina:2020xhk,Baldes:2020nuv,Gondolo:2020uqv,Bernal:2020kse,Bernal:2020bjf}   
or provide seeds for the supermassive black holes observed at the centers of galaxies~\cite{Bean:2002kx,Kawasaki:2012kn}.

Theoretical studies of black hole thermodynamics suggest that black holes have a temperature that is inversely proportional to the black hole's mass, $T_\text{PBH} \sim 1/M_\text{PBH}$, and that they emit particles as Hawking radiation~\cite{Hawking:1975vcx}.  As a black hole radiates, it loses mass and heats up, leading to a runaway evaporation process, known as an explosion~\cite{Hawking:1974rv}.  While the stellar mass and supermassive black holes observed so far have a very low temperature and will not explode any time soon, Schwarzschild PBHs with an initial mass around $6\times 10^{14}$\,g could be exploding today.

Observation of an exploding black hole would be transformative in at least three ways: it would provide strong evidence for the existence of PBHs, the first evidence of Hawking radiation and definitive information about the particles present in nature.  While the first two points are clear, the third is less obvious. 
 Theoretically, it is expected that a black hole will radiate all fundamental particles with a mass below the black hole temperature, regardless of their non-gravitational interactions.  This means that the gamma ray signal from an exploding black hole encodes information about all fundamental particles with masses less than the Planck scale~\cite{Baker:2021btk,Baker:2022rkn,Boluna:2023jlo,Federico:2024fyt}.  Observation of a PBH explosion would go far beyond other known ways of searching for new particles, such as particle colliders.

Current TeV-scale gamma ray observatories such as HAWC and LHAASO are capable of observing exploding black holes in the vicinity of the Earth (to a distance of around 0.1 parsecs).\footnote{Note that if an observation is made, follow up experiments looking for the absence of a lower energy afterglow would be required to determine that the gamma ray burst most likely originated from a PBH explosion.} The HAWC collaboration has not yet observed an event and has set the strongest direct upper limit on the local burst rate, $\dot{n}_{\rm PBH}=3400\,\text{pc}^{-3}\text{yr}^{-1}$~\cite{HAWC:2019wla}, while the LHAASO collaboration has the capacity to improve the limit to $\dot{n}_{\rm PBH}=1200\,\text{pc}^{-3}\text{yr}^{-1}$~\cite{Yang:2024vij} in the near future. However, if these experiments hope to observe an exploding black hole, it is expected that the black hole is part of a population of PBHs with a temperature around $100\,\text{MeV}$.  This population will contribute to the Extragalactic Gamma-Ray Background (EGRB)~\cite{Carr:2020gox}, which under motivated assumptions limits the local burst rate to less than $\dot{n}_{\rm PBH}\sim0.01\,\text{pc}^{-3}\text{yr}^{-1}$~\cite{Boluna:2023jlo}, far beyond the reach of direct observation.

However, these direct limits and indirect constraints assume that the PBHs are Schwarzschild (that is, they have no electric charge or angular momentum).  In this letter, we introduce a new mechanism which we demonstrate in an example scenario featuring a new $u(1)$ gauge symmetry, a heavy dark electron, and where PBHs are formed with some dark charge (similar to scenarios considered in Ref.~\cite{Bai:2019zcd}, to accommodate light PBH dark matter, and in Ref.~\cite{Barbosa:2025uau}).  
We find that these PBHs can undergo a period of quasi-extremality, which weakens the indirect constraints and provides an enhancement of the possible local burst rate. We find that in this scenario current gamma ray observatories could soon observe a black hole explosion.

\section{The Argument Against Observing an Exploding PBH}

Many different observables can be used to constrain the presence of PBHs over a wide range of masses (see~\cite{Green:2020jor, Carr:2021bzv} and references therein for a comprehensive overview).  It is usually assumed that all PBHs are Schwarzschild and that they only evaporate into Standard Model particles.  Under these assumptions PBHs with an initial mass of around $M_{c}^{S} \simeq 5.6 \times 10^{14}$\,g would explode today. For these masses, the strongest bound is due to photons emitted as prompt Hawking radiation or as secondary radiation from other radiated particles (due to hadronization, final state radiation, $e^{+}e^{-}$ annihilation, etc).  This photon emission would contribute to the observed galactic and extragalactic X-ray and gamma-ray fluxes.  The dominant constraint is from the EGRB and the maximum allowed abundance is $f_{\rm PBH} \sim 10^{-10}$~\cite{Carr:2009jm,DelaTorreLuque:2024qms}, where $f_\text{PBH} = \Omega_\text{PBH}/\Omega_\text{DM}$ and where $\Omega_\text{PBH}$ and $\Omega_\text{DM}$ are the energy densities in PBHs (at the time of formation) and dark matter, respectively.   For a monochromatic mass function (where all PBHs have the same mass) this constraint is shown in the grey region in \cref{fig:indirect_bounds}~\cite{Carr:2009jm}.  When considering realistic scenarios with log-normal mass distributions and a width $\sigma$ between 0.05 and 0.5~\cite{Boluna:2023jlo}, this leads to a local burst rate of $\dot{n}_{\rm PBH}\lesssim 0.01\,\text{pc}^{-3}\text{yr}^{-1}$. Since current telescopes can only directly observe an exploding PBH within $\mathcal{O}(0.1)$\,pc of Earth, this means that we expect to observe no more than one event every $\mathcal{O}(10^{5})$ years.  
This is essentially the argument against observing an exploding PBH.

Before we go beyond the usual assumptions, we note that although the small value of $f_{\rm PBH}$ leads to low burst rates in the above scenario, this is not always the case. While the PBH abundance is typically quoted in terms of $f_{\rm PBH}$, a given $f_{\rm PBH}$ can be in the form of many light PBHs or fewer heavier PBHs.  The burst rate depends on the PBH number density in the vicinity of Earth,
\begin{equation}
    n_{\rm PBH} = f_{\rm PBH}\frac{\rho_\text{DM}}{M_{\rm PBH}},
    \label{Ndensity}
\end{equation} 
where we take $\rho_{DM} = 0.4\,\text{GeV} \,\text{c}^{-2} \text{cm}^{-3}$ as the local dark matter density \cite{Benito:2019ngh} and $M_{\rm PBH}$ is the typical mass of a PBH.  For the same $f_\text{PBH}$, a higher abundance of lighter PBHs will give a larger burst rate. While this observation is not very useful in the standard scenario, since PBHs lighter than $M_{c}^{S}$ will have already exploded, we now explore a scenario where light PBHs are long-lived and could explode today.

\section{Quasi-extremal Black Holes with Physics Beyond the Standard Model}

In this section we introduce a mechanism that leads to long-lived PBHs. While the mechanism could work in a wide variety of models, as a toy example we extend the Standard Model with a new dark $u(1)$ gauge symmetry and a dark electron. The new Lagrangian terms are
\begin{align}
    \mathcal{L} 
    \supset
    - \frac{1}{16\pi}F'^{\mu\nu}F'_{\mu\nu}
    +
    \overline{\ell_D} (i\slashed{D} - m_D)\ell_D
    \,,
\end{align}
where $F'_{\mu\nu} = \partial_\mu A'_\nu - \partial_\nu A'_\mu$ is the field strength tensor of the dark photon, $D_\mu = \partial_\mu + i e_D A'_\mu$ is a covariant derivative,\footnote{Note that we use units where $c = \hbar = 4\pi\epsilon_0 = 4\pi\epsilon_0' = 1$ (where $\epsilon_0'$ is the dark vacuum permittivity) and $M_\text{Pl}=G^{-1/2} = 2.18 \times 10^{-5}\,\text{g} = 1.22 \times 10^{19}\,\text{GeV}$.} and $m_D$ and $e_D$ are the mass and dark charge of the dark electron, respectively.  We assume that the kinetic mixing term $\epsilon F_{\mu\nu}F'^{\mu\nu}$ is negligible and that the dark photon is massless. 

We now describe the essential features of the mechanism.  We first imagine that a Reissner–Nordstr\"om (RN) PBH is formed in the early universe, with an initial non-zero dark charge $Q_D^i$.\footnote{In this work we will take $Q_D^{*i} \sim 0.01$. 
 PBHs of a mass $10^{11}$ to $10^{13}$\,g would typically have $Q_D^{*i} \sim 10^{-4}$ due to random charge fluctuations during formation~\cite{Bai:2019zcd}, so we assume a mild asymmetry generation mechanism.}  Neglecting accretion, this PBH will start to lose mass due to Hawking radiation.  If the dark electrons are much heavier than the PBH's temperature, $Q_D$ will not change and the PBH will become quasi-extremal, $Q^*_D \equiv Q_D M_{\rm Pl}/M_\text{PBH} \rightarrow 1$. 
The temperature of an RN black hole is given by~\cite{Page:1976df}
\begin{align}
    T_{\rm PBH} = \frac{M_{\rm Pl}^{2}}{2 \pi M_{\rm PBH}} \frac{\sqrt{1 - (Q_D^{*})^{2}}}{\left( 1 + \sqrt{1 - (Q_D^{*})^{2}} \right)^{2}}.
    \label{eq:Tbh}
\end{align}
For fixed $M_{\rm PBH}$, as $Q_{D}^{*} \to 1$ the temperature reduces and Hawking emission becomes heavily suppressed. The PBH will then remain quasi-extremal with a constant dark charge and a roughly constant mass for a long period of time. However, since the black hole is not exactly extremal, it will very slowly lose mass via the emission of light and massless SM particles. As it loses mass, the black hole shrinks and the dark electromagnetic field near the event horizon gets stronger. 
 Eventually, it will become so strong that dark electron-positron pairs form via the (dark) Schwinger effect~\cite{Schwinger:1951nm}.  This will quickly discharge the black hole~\cite{PhysRevLett.33.558}, increasing its temperature and leading to a Schwarzschild-like explosion.  In this scenario, PBHs that explode today will be lighter at formation than in the standard Schwarzschild case due to the long period of suppressed emission, enhancing the burst rate. The lower temperature and suppressed grey body factors also reduce the contributions to the EGRB, further weakening the indirect constraints.
 
Quantitatively, in the limits where the black hole radius is much larger than the (reduced) Compton wavelength of the dark electron, $M_\text{PBH}/M_{\rm Pl}^{2} \gg 1/m_D$, the potential energy gained by a dark electron repelled from the black hole horizon to infinity is much greater than its rest mass, $Q_D e_D/r_+ \gg m_D$, and where the black hole is in the very slow discharge regime~\cite{PhysRevLett.33.558},
\begin{align}
\label{eq:limit-2}
    M_{\rm PBH} \gg M_{d} \equiv \frac{e_D M_\text{Pl}^3}{\pi m_D^2}
    \,,
\end{align}
the black hole mass and charge evolution are governed by~\cite{PhysRevD.41.1142,Cohen:2008wz}
\begin{align}
\label{eq:m-evolution}
    \frac{dM_{\rm PBH}}{dt} &= -\frac{\alpha(M_{\rm PBH}, Q_D)}{M^{2}_{\rm PBH}}+\frac{Q_D}{r_{+}}\frac{dQ_D}{dt}\,,
    \\
    \frac{dQ_D}{dt} &\simeq -\frac{e_D^{4}}{2 \pi^{3} m_D^2} \frac{Q_D^{3}}{r_{+}^{3}} \text{exp} \left(\frac{-\pi r_{+}^{2} m_D^2}{Q_D\,e_D}\right)\,,
\label{eq:q-evolution}
\end{align}
where
\begin{align}
    r_{+}=\frac{M_{\rm PBH}}{M_{\rm Pl}^{2}}\left(1 + \sqrt{1 - (Q_D^\ast)^{2}
    }\right)
\end{align}
is the outer horizon radius of a RN black hole.  The Page function $\alpha(M_{\rm PBH}, Q_D)$ is
\begin{align}
    \alpha(M_{\rm PBH}, Q_D) &= M_{\rm PBH}^{2} \sum_{i} \int_{0}^{\infty}\frac{d^{2}N_{i}}{dEdt}E \, dE \,, 
    \\
    \text{with} \quad \frac{d^{2}N_{i}}{dEdt} &= \frac{n_{\rm dof}^{i}\Gamma^{i}(M_{\rm PBH},E,Q_D)}{2\pi(e^{E/T_{\rm PBH}} \pm 1)} \,,
    \label{eq:d2NdEdt}
\end{align}
where $E$ is the energy of the particle emitted via Hawking radiation and the sum is over all emitted fermions (+) and bosons ($-$) in the SM and dark sector. The number of degrees of freedom of the $i$-th particle is given by $n_{\rm dof}^{i}$ and we use the greybody factors, $\Gamma^i(M_{\rm PBH},E,Q_D^*)$, provided by \texttt{BlackHawk}~\cite{Arbey:2019mbc,Arbey:2021mbl} up to the maximum value of the charge parameter for which they are available, $Q_D^{*}=0.999$. These greybody factors are given for the standard electric charge, but are also valid for a dark electric charge.  For $Q_D^{*}>0.999$, we use the greybody factors at $Q_D^{*}=0.999$ and retain only the dependence on $Q_D^{*}$ through the temperature in the argument of the exponential in \cref{eq:d2NdEdt}. Due to the exponential suppression, this approach accurately captures the leading dependence on the charge parameter in this regime.

\begin{figure}
    \centering
    \includegraphics[height=0.3\textwidth]{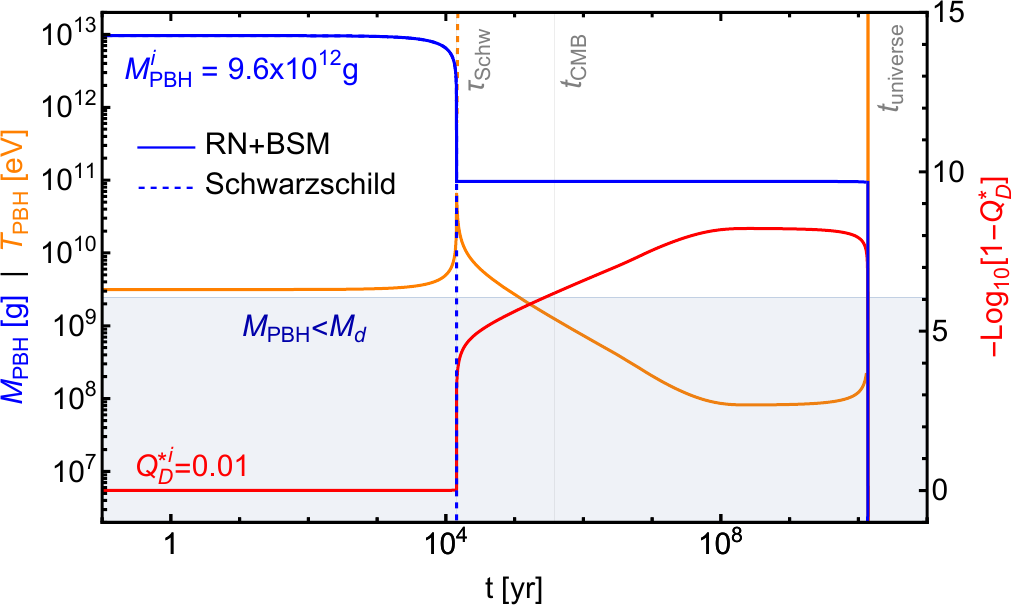}
    \caption{PBH mass (blue), charge (red) and temperature (orange) evolution for $m_D=10^{10}\,$GeV and $e_D=10^{-3} e_{\rm SM}$. We also show $\tau_{\rm Schw}$ (the lifetime of a  Schwarzschild PBH of the same initial mass), $t_{\rm CMB}$ (the time of recombination) and $t_{\rm universe}$ (the age of the universe today).
    }
    \label{fig:pbh-evolution}
\end{figure}

In \cref{fig:pbh-evolution} we show the time evolution of the PBH mass (blue), charge parameter (red) and temperature (orange) of a Schwarzschild (dotted) and RN (solid) PBH with initial mass $M^{i}_{\rm PBH} = 9.6 \times 10^{12}$\,g and for the RN PBH an initial charge parameter $Q_D^{*i} = 0.01$, for $m_D=10^{10}\,$GeV and $e_D=10^{-3} e_{\rm SM}$.  We see that a Schwarzschild PBH of this mass would fully evaporate around $10^4$ years after the big bang.  The RN black hole starts to lose mass in the same way but the explosion is halted by the increasing charge parameter, which dramatically decreases the PBH's temperature and heavily suppresses Hawking radiation.  The quasi-extremal PBH is then essentially stable until the dark Schwinger effect becomes efficient, at around $10^{10}$\,years, at which point the PBH rapidly discharges (passing from a very slow discharge regime to a very rapid one, the superradiant discharge regime~\cite{PhysRevLett.33.558}).  The PBH then follows a Schwarzschild-like explosion.

\begin{figure}
    \centering
    \includegraphics[height=0.3\textwidth]{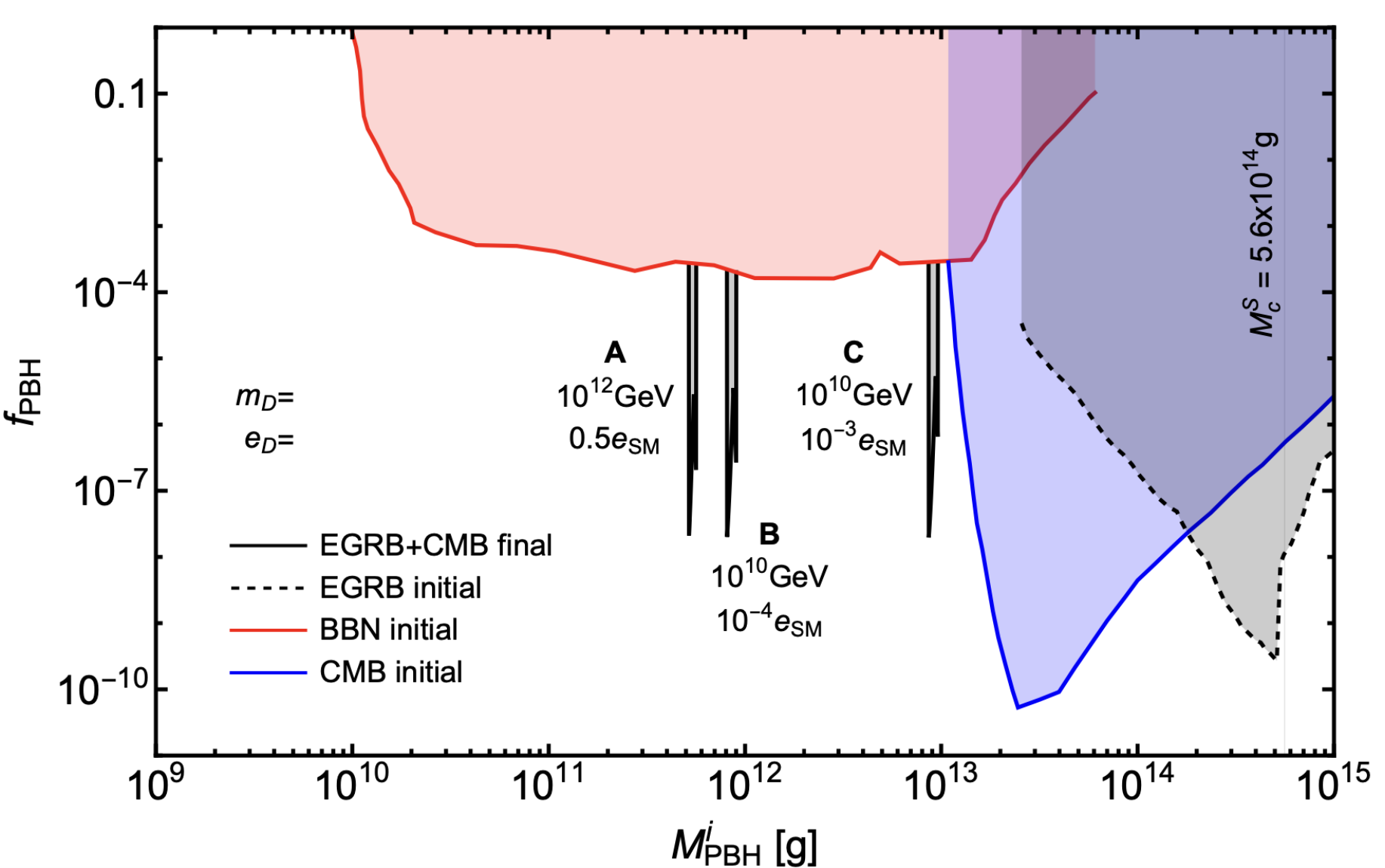}
    \caption{Indirect bounds on $f_{\rm PBH}$ from BBN (red), CMB (blue) and the EGRB (grey) during the initial evaporation phase and from both the CMB and EGRB (black) during the final explosion for three different benchmark points (A, B and C). 
    }
    \label{fig:indirect_bounds}
\end{figure}

For a population of RN PBHs of the same initial mass and charge parameter, significant radiation will be emitted when the PBHs are Schwarzschild-like, both before becoming quasi-extremal and during the final explosion.  The initial emission is constrained by Big Bang Nucleosynthesis (BBN), as energy injection can alter the final abundances of light elements, and potentially by the Cosmic Microwave Background (CMB), as it may affect anisotropies in the CMB.  We show these as red and blue regions in \cref{fig:indirect_bounds}, respectively.  
For the mass range of interest ($10^{10}$ to $10^{13}$\,g) the PBHs are Schwarzschild-like during BBN, so the BBN constraints are not significantly modified, but the PBHs are quasi-extremal by the time of recombination.
The emission from the final explosions is constrained by both the CMB and the EGRB. In \cref{fig:indirect_bounds} we show these constraints in black for several dark sector parameter choices.  
To set the EGRB limits we use \texttt{PYTHIA}~\cite{Bierlich:2022pfr} and \texttt{Hazma}~\cite{Coogan:2019qpu} in their corresponding energy ranges of validity to compute the secondary photon flux~\cite{Coogan:2020tuf} and limit the total photon flux by available EGRB data, which we fit with a power law ($d\phi/dE_{\gamma} \propto E_{\gamma}^{-\alpha}$ with $\alpha=2.6$ in the energy range 1\,MeV to $10^6$\,GeV, based on observations from $\sim 1$\,MeV to $\sim 1$ TeV and consistent with the limits up to $10^6$\,GeV~\cite{Fermi-LAT:2014ryh,HAWC:2022uka}). This constraint gives a sharper spike than in the Schwarzschild case as slightly heavier PBHs are still quasi-extremal today with suppressed emission.  
Slightly lighter black holes exploded in the past and their abundance can be constrained by studying the effect of the energy injection on the CMB anisotropies. We estimate the CMB bound following the procedure described in reference~\cite{Carr:2009jm}, adapted to our scenario.\footnote{This provides a good estimate of the bound while avoiding a full analysis of the ionization history of the universe~\cite{Acharya:2020jbv}.} Once again we note a sharper spike than in the Schwarzschild case since slightly lighter PBHs became quasi-extremal before recombination. 
For different points in the dark sector parameter space, the initial mass of a PBH that would be exploding today (the critical mass, $M_{\rm c}$) roughly scales as $\sim e_D/m_D^{2}$, as long as $M_{\rm c}$ is well below $M_{c}^{S}$.

Finally, we consider a more realistic situation where the PBHs are formed with a log-normal mass distribution with a plausible width of $\sigma_M=0.3$~\cite{Gow:2020bzo,Gow:2020cou} and pivot mass $\bar{M}$, and track their evolution using \cref{eq:m-evolution,eq:q-evolution}.  We still assume all PBHs are formed with the same initial charge parameter $Q_D^{*i}$, but discuss this assumption below. To translate the monochromatic observational constraints shown in \cref{fig:indirect_bounds} to constraints on a log-normal mass distribution we follow the procedure outlined in Ref.~\cite{Carr:2017jsz}. 

\begin{figure}
    \centering
    \includegraphics[width=0.5\textwidth]{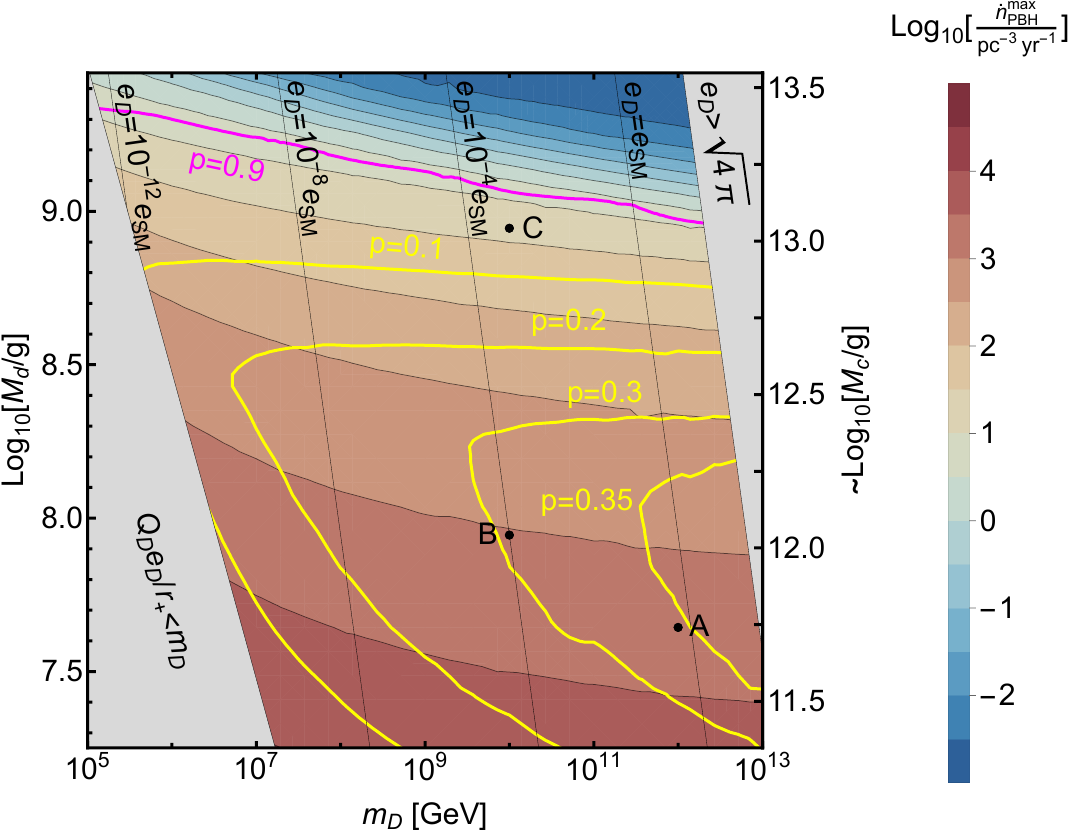}
    \caption{Maximum burst rates consistent with the CMB and EGRB constraints for a log-normal mass distribution of PBHs with $\sigma_M = 0.3$ and $Q_D^{*i}=0.01$, along with the corresponding probability of an observation at HAWC (yellow) and LHAASO (magenta) with 10 years of data. The $y$-axis shows both the discharge mass (left) and the approximate critical mass (right).  In the upper right corner perturbativity does not apply while on the far left the condition $Q_D e_D/r_+ > m_D$ is not satisfied.  We also indicate the positions of the benchmark points A, B and C.}
    \label{fig:burst_rate}
\end{figure}

In \cref{fig:burst_rate} we show the maximum local burst rate allowed by the indirect bounds on $f_{\rm PBH}$, $\dot{n}_\text{PBH}^\text{max}$, on the dark sector parameter space (following Ref.~\cite{Boluna:2023jlo}), along with the probability of observation at HAWC and LHAASO over 10 years of observation (yellow and magenta solid lines) assuming a population of PBHs which saturates the CMB and EGRB limits. 
We see that while the standard limit of $\dot{n}_{\rm PBH} \lesssim 0.01\,\text{pc}^{-3}\text{yr}^{-1}$ applies in the upper region of the parameter space (where the Schwinger effect discharges the black hole a long time before the final explosion), the local burst rate can be as large as $\dot{n}_{\rm PBH} \sim 10^4\,\text{pc}^{-3}\text{yr}^{-1}$ in the lower region. As we have discussed, quasi-extremal PBHs have a suppressed Hawking emission, so PBHs that are lighter when they discharge emit fewer photons in their final burst.  This means that they are less constrained by the CMB and EGRB constraints, but are also less visible to experiments like HAWC. There is a trade-off between the burst rate and the burst duration, with the highest probability of observation occurring around $\dot{n}_{\rm PBH}^\text{max} \sim 1000\,\text{pc}^{-3}\text{yr}^{-1}$. On the right-hand side of \cref{fig:burst_rate} this trade-off is optimised. Furthermore, due to this effect, we estimate that the direct HAWC limit~\cite{HAWC:2019wla} on RN PBHs is weaker than the indirect constraints.
In the grey region on the right-hand side of \cref{fig:burst_rate} our perturbative calculations are no longer valid, while on the left-hand side an assumption in our evolution equations is violated.
We have checked that wider log-normal mass distributions do not significantly change the results shown in \cref{fig:burst_rate}. Indeed, as long as there is not a significant population above $M_{\rm PBH} \sim 2\times 10^{13}$g, the parameter space is quite unconstrained.

\section{Conclusions}

In this letter we have for the first time outlined a plausible scenario where current experiments could observe an exploding PBH.  
Using the example scenario of a new dark $u(1)$ gauge symmetry we have shown that PBHs may undergo a period of quasi-extremality, suppressing their Hawking radiation and allowing lighter PBHs to survive until today.  This heavily weakens the indirect constraints on the burst rate and boosts the possible observation probabilities.  
Beyond this simple example, the same general mechanism could also occur in more sophisticated models and potentially
in situations where PBHs approach extremality by increasing their spin~\cite{deFreitasPacheco:2023hpb}, increasing their magnetic charge~\cite{Bai:2020spd,Kobayashi:2023ryr} or in models involving extra dimensions~\cite{Friedlander:2022ttk}. 
Exploding PBHs could provide transformative insights into our universe: they would provide evidence for the existence of PBHs, evidence of Hawking radiation and give definitive information on the fundamental particles present in nature. We should ensure that we are well prepared to make the most of a nearby exploding PBH in the near future.

\appendix


\section{The Probability of Observation}

In this paper we describe a novel mechanism that enhances the possible local burst rate by several orders of magnitude. Here we discuss how this increases the probability of observing an exploding PBH. Intuitively, as the burst rate increases, more explosions will occur in the local vicinity of Earth, making it more likely to detect an explosion. However, the higher allowed burst rate comes with a shorter burst duration. This results in fewer high-energy photons being emitted, making detection more challenging. 

To calculate the probability that at least one black hole explosion will be observed by HAWC or LHAASO in 10 years of observation, we compute the distance at which the observatories would be able to observe 10 photons with energy $E_{\gamma} \geq 10^{3}$\,GeV~\cite{Baker:2025-forthcoming}.  We assume that the exploding PBH behaves exactly like a Schwarzschild black hole once it reaches $M_{\rm PBH}=M_d$~\cite{PhysRevLett.33.558}. This assumption is well justified since at this point the PBH will have lost nearly all of its charge via the Schwinger effect, resulting in a charge parameter close to zero. 
Using this distance, we compute the volume of space where HAWC or LHAASO could observe an exploding black hole in our scenario, $V_{\rm search}$. The average number of exploding PBH observations in a given time $T_\text{obs}$, $\bar{N}$, is then
\begin{equation}
    \bar{N}
    =
    \dot{n}_{\rm PBH}
    V_\text{search}
    T_\text{obs}
    \,.
\end{equation}
The probability density function for the number of observations follows a Poisson distribution with mean $\bar{N}$, so the probability to observe at least one exploding PBH is
\begin{align}
    \text{Pr}(N_\text{obs}\geq1)
    &=
    1-\text{Pr}(N_\text{obs}=0) = 1-\exp(-\bar{N})
    \,.
\end{align}

Since HAWC is relatively insensitive to the energy of incoming gamma rays, we can also estimate the direct limit placed on RN PBH explosions by rescaling HAWC's direct limit on Schwarzschild PBH explosions, ($3400\,\text{pc}^{-3}\text{yr}^{-1}$)~\cite{HAWC:2019wla}, using the ratios of search volumes in the two scenarios. In the parameter space under consideration the resulting limit is always weaker than that from indirect constraints. While this estimate is encouraging, a dedicated analysis would be needed to obtain a more accurate direct limit.

\section{The Initial Dark Charge of the PBHs}

In the results we presented in \cref{fig:burst_rate}, we assumed a formation mechanism which produces PBHs with a small but significant charge ($Q^{*i}_D=0.01$). We have checked that the maximum allowed burst rate does not depend strongly on the initial dark charge assumed, as long as the dark electron mass and coupling are adjusted appropriately. 
 Explicit mechanisms for charged PBH formation are not well-explored in the literature, since for the Standard Model electromagnetic charge a PBH would quickly neutralize via interactions with surrounding electrons or positrons or via evaporation into charged particles~\cite{PhysRevD.16.2402}. However, in the scenario of interest, there are several reasons to think that neutralization would be suppressed. 
 When the dark electron is sufficiently massive, Hawking evaporation is suppressed and the PBH cannot neutralize via emission.   PBH neutralization via accretion may also be suppressed for several reasons. 
 Firstly, the light black holes we consider are very small, suppressing accretion (which typically scales as their area).  Secondly, in-medium effects similar to those discussed in Ref.~\cite{Alonso-Monsalve:2023brx} can screen the PBH.   
 Finally, when the temperature of the dark sector plasma, secluded from the Standard Model, drops below the dark electron mass after PBH formation, dark electrons and positrons will annihilate away from the plasma.  This suppresses their number density and hence the neutralization rate~\cite{Bai:2019zcd}.
 Alternatively, one could invoke an asymmetric dark sector~\cite{Bramante:2024pyc,Affleck:1984fy}, possibly motivated by the baryon asymmetry in the visible sector, which would naturally lead to significantly charged PBHs.

We also assumed that all PBHs are formed with exactly the same charge parameter $Q_D^{*i}$. However, in reality, one would expect an extended distribution in charge as well as in mass.  We here estimate the effect of an extended charge distribution (at a single choice of dark electron mass and coupling) on the resulting burst rates. In the Schwarzschild case, given an initial mass distribution, $\psi_{i}$, the burst rate will depend on the value of the mass function evaluated at $M_{c}^S$~\cite{Boluna:2023jlo},
\begin{equation}
    \dot{n}_{\rm PBH} \propto \frac{\rho_{\rm DM}}{t_{U}} \psi_{i}({M_{c}^S)}
    \,,
    \label{eq:schwarzschil-rate}
\end{equation}
where $t_{U}$ is the age of the universe.
When considering an extended charge distribution, there will not be a single value for $M_{c}$, instead there will be a continuum of initial PBH configurations $Q_{c}(M)$ on the space $\{M_{i},Q_{i}\}$ which corresponds to PBHs exploding today. Using a log-normal mass distribution and modeling the charge distribution with a Gaussian function with mean $\bar{Q}$ and standard deviation $\sigma_Q$, the burst rate will be
\begin{equation}
\begin{split}
    \dot{n}_{\rm PBH} \propto & \frac{\rho_{\rm DM}}{t_{U}} \int dQ \int \frac{dM}{\sqrt{2\pi} \sigma_{M} M^{2}} \text{exp} \left(-\frac{\text{log}(M/\bar{M})^{2}}{2 \sigma_{M}^{2}} \right) \\
    &\times \frac{1}{\sqrt{2\pi} \sigma_{Q}} \text{exp} \left(-\frac{(Q-\bar{Q})^{2}}{2 \sigma_{Q}^{2}} \right) \delta\left(1-\frac{Q_{c}(M)}{Q}\right)
    \,,
\end{split}
\label{eq:extended-charge-distribution-rate}
\end{equation}
where the Dirac delta function $\delta(1-Q_{c}(M)/Q)$ restricts the integral to the line in the space $\{M_{i},Q_{i}\}$ such that the lifetime of the black hole is equal to the age of the universe, $\tau_{\rm PBH}=t_{U}$. Note that in the limit $\sigma_{Q}\rightarrow 0$ the Gaussian distribution becomes a Dirac delta function (leading to $M_{c} \delta(M-M_{c})$) and we recover the expression in \cref{eq:schwarzschil-rate}. From \cref{eq:extended-charge-distribution-rate} we find that as long as $Q_{c}$ is roughly in the interval $(\bar{Q} - 2\sigma_{Q},\bar{Q} + 2\sigma_{Q})$, the final burst rate is only modified by an $\mathcal{O}(1)$ factor for both narrow and wide charge distributions (in fact for $\sigma_{Q} \lesssim 6 Q_{c}$).

Finally, we briefly consider going beyond log-normal mass distributions.  In particular, the popular critical collapse formation mechanism may produce a population of PBHs with an enhanced tail at small masses (if, e.g., the peak in the power spectrum is relatively narrow)~\cite{Gow:2020cou}.  Although these small-mass PBHs would only constitute a fraction of the PBHs, it is plausible that they could form with a large $Q_D^*$~\cite{Alonso-Monsalve:2023brx} which would increase their lifetime.  These light PBHs could then contribute to the EGRB, CMB and BBN constraints, reducing the maximum allowed PBH abundance, and they could also contribute to the observable burst rate today.  In the scenario we discuss, the lifetime of a dark charged PBH is very sensitive to its initial conditions (e.g., in \cref{fig:indirect_bounds} we see that PBHs that can impact the CMB have almost the same initial mass as PBHs exploding today).  Furthermore, we note that the mechanism we discuss relaxes the indirect bounds and increases the burst rate primarily by reducing the initial mass of the PBHs, c.f.~\cref{Ndensity}.  This means that as long as the initial mass-charge distribution does not have very sharp features, these light PBHs will contribute more to the burst rate than to the indirect constraints.  We therefore expect our main conclusions to be conservative for mass distributions with enhanced small-mass tails.  However, as the precise impact depends on the whole evolution of this population, a more detailed study is required to confirm this conclusion.

\section{A Note on Units}

Since this work straddles particle physics (where the convention $\epsilon_0 = 1$ is typically used) and general relativity (which often uses $4\pi\epsilon_0 = 1$), a note on units is in order.  In SI units the SM QED Lagrangian density is~\cite{Lv:2015zpa}
\begin{align}
    \mathcal{L}
    &=
    -\frac{1}{4\mu_0}
    F^{\mu\nu} F_{\mu\nu}
    +\bar{\psi}(i\hbar c \slashed{D} - mc^2)\psi
    \,,
\end{align}
where $D_\mu = \partial_\mu +\frac{ie}{\hbar} A_\mu$ and the fine-structure constant $\alpha = e^2/4\pi\epsilon_0 \hbar c$.

If the convention $c = \hbar = \epsilon_0 = 1$ is used then $e = \sqrt{4\pi\alpha}$ and the kinetic term of the photon becomes $-\frac{1}{4}F^{\mu\nu} F_{\mu\nu}$ (since $\epsilon_0 \mu_0 = 1/c^2$).  In this convention the perturbativity limit is (usually taken to be) $e<4\pi$.

In the convention $c = \hbar = 4\pi\epsilon_0 = 1$ we have $e = \sqrt{\alpha}$ and the kinetic term of the photon becomes $-\frac{1}{16\pi}F^{\mu\nu} F_{\mu\nu}$.  Since the kinetic term has a new normalisation, the Feynman rules for photon propagators and external states obtain factors of $4\pi$ so that matrix elements do not depend on the convention when written in terms of $\alpha$.  In this convention, the perturbativity limit becomes $e<\sqrt{4\pi}$.


\bibliography{biblio}

\end{document}